# Quantum sensing of radiofrequency fields using cold Rydberg atoms in a super-molasses trap

Romain Granier[1], Anthony El Bekai[1], Miguel Angel Cifuentes Marín[1], Cédric Blanchard[1], Nassim Zahzam[1], Yannick Bidel[1], Alexandre Bresson[1], Vilius Atkočius[2], Chester Camm[2], Florence Concepcion[2], Konstantinos Karakostas[2], Matt Himsworth[2], Alexander Jantzen[2], Alexis Bonnin*[1] and Sylvain Schwartz*[1]

[1] DPHY, ONERA, Université Paris-Saclay, 91120 Palaiseau, France
[2] Aquark Technologies Ltd, Eastleigh, S050 4SR, UK
* Corresponding authors: alexis.bonnin@onera.fr, sylvain.schwartz@onera.fr

We demonstrate the quantum sensing of radiofrequency fields based on cold Rydberg atoms in a super-molasses trap, without the need for magnetic coils. Our approach combines the metrological advantages of cold atoms with a metal-free dielectric sensor head minimizing perturbations to the electromagnetic environment, a feature that was previously restricted to vapor-cell-based devices. Moreover, the absence of inductive loads allows to rapidly alternate between the cooling phase and the Rydberg excitation, leading to trap-loss spectroscopy signals one order of magnitude narrower than for conventional magneto-optical traps. This enables self-calibrated microwave power measurements with an unprecedented dynamic range of 43 dB, opening the door to new perspectives of applications in calibration measurements. We also report a scale factor linearity better than 1%, the absence of drifts over several tens of minutes leading to a 3 $\mu$V/cm resolution, and the possibility to retrieve the ellipticity of the applied microwave field. By demonstrating a cold-atom metrological platform in a compact dielectric sensor head, this work paves the way for new applications in the field of radiofrequency measurements with Rydberg atoms, and in other fields of quantum sensing based on cold atoms such as magnetometry, gravimetry or inertial navigation.

## INTRODUCTION

Rydberg atoms (1) are atoms excited to electronic states with a high value of their principal quantum number $n$. The large transition dipole moments between neighboring Rydberg states, scaling as $n^2$, make them very sensitive to external electric fields over a wide range of frequencies, extending from the MHz to the THz domains. This has motivated a very active research effort to develop a new class of radio-frequency (RF) quantum sensors based on electromagnetically-induced transparency (EIT) involving Rydberg states in thermal vapor cells (2-5). In addition to their intrinsically high sensitivity, Rydberg RF quantum sensors offer several advantages compared to conventional antenna-based technologies. This includes the possibility to build a fully dielectric sensor head to minimize perturbations of the surrounding electromagnetic environment, typically using a fiber-coupled vapor cell (6,7). Besides, Rydberg RF quantum sensors can perform self-calibrated measurements of a resonant RF field interacting with the atoms, by measuring the frequency of the Autler-Townes splitting of the EIT peak. In this situation, the measured frequency is proportional to the amplitude of the RF field with a scale factor determined by fundamental constants and intrinsic atomic properties only, making the measurement calibration-free and directly traceable to the international system of units (SI) (8).

The minimal SI-traceable amplitude of the RF field that can be resolved using Autler-Townes splitting is set by the linewidth of the EIT peak, which is on the order of a few MHz in an alkali-vapor cell and depends on various parameters including laser power, lifetime of the intermediate state, Doppler effect, transit time broadening and collisions (9,10). A first approach to go beyond this limit is the use of a three-photon scheme in a vapor cell, instead of the usual two-photon scheme, where the sum of the wavevectors is close to zero, which reduces the contribution of the Doppler effect to the EIT linewidth (11-13). For cesium, this can be achieved to a very good extent with collinear beams as reported in reference (12), where a sub-200-kHz EIT linewidth (full width at half maximum, FWHM) was achieved, with a factor on the order of 50 between the lowest and the highest measured RF power, corresponding to an SI-traceable dynamic range of 17 dB. Another approach is the use of cold atoms (14-20), where Doppler effect and transit time broadening can be strongly reduced by laser cooling techniques. For example, in reference (15), a Rydberg-EIT peak with a linewidth of about 1.1 MHz is shown, with an SI-traceable dynamic range on the order of 28 dB. The strong reduction of Doppler effect in cold atoms makes it also possible to use a direct 2-photon coupling between the ground and the Rydberg states with a large detuning from the intermediate state (19), the linewidth being in this case ultimately limited by the lifetime of the Rydberg state, on the order of 1.5 kHz for $61S_{1/2}$ in $^{87}Rb$ (21). Furthermore, cold atoms allow a local, sub-millimeter, measurement of the electromagnetic field in all three dimensions, mitigating the broadening of the Autler-Townes peaks for increasing values of the RF power induced by spatial inhomogeneities (22). Finally, cold atoms are well-isolated from their environment, making them natural candidates for precision measurements and metrology. However, all the experiments mentioned previously start with a magneto-optical trap, which requires at least one pair of coils around the atoms, and compromises the possibility of building a fully dielectric sensor head. Furthermore, rapid modulation of the magnetic field created by inductive coils remains a significant technical challenge (23,24).

In this paper, we report the measurement of RF fields based on cold rubidium Rydberg atoms without the need of magnetic coils. Our approach is based on a super-molasses trap (SMT), similar to the one recently described in reference (25). It typically produces a cloud of $10^7$ atoms at a temperature of 300 µK with an all-optical, fully dielectric arrangement around the atoms. Owing to the reduced Doppler effect, we can perform a direct 2-photon transition between the ground and the Rydberg states in a trap-loss measurement, similar to reference (19). Moreover, we take advantage of the absence of slow inductive loads to rapidly alternate between the cooling and the Rydberg beams, suppressing broadening effects, lightshifts and ground state Autler-Townes splitting usually observed in trap-loss spectroscopy from a magneto-optical trap (19, 26, 28). This results in high-contrast spectral features as narrow as 1.5 MHz (FWHM), and Autler-Townes splitting as high as 223 MHz, corresponding to more than 43 dB of SI-traceable RF power dynamic range. We also show a sub-percent linearity over a 32 dB RF power range, and a resolution of 3 $\mu$V/cm after a few minutes of integration through Allan variance measurements. By combining the advantages of cold atoms with the possibility of an all-optical, fully dielectric sensor head, this work pushes the limits of practical RF quantum sensors based on Rydberg atoms, and opens new avenues for applications of cold atom systems in various areas of quantum technologies.

## RESULTS

### Trap-loss spectroscopy in the absence of cooling light

We start with an all-optical source of 87 Rubidium atoms in a super-molasses trap (SMT), similar to the one described in (25). The dielectric sensor head consists of a glass cell pumped by a vacuum system located far away from the measurement area, and surrounded by a ceramic resin support on which the required free-space optics are mounted (Fig. 1A). With this system, we obtain an atomic cloud of about $10^7$ atoms at $300\ \mu K$ in a volume of $0.5\text{ mm} \times 0.2\text{ mm} \times 0.2\text{ mm}$ without using any magnetic coils. To couple the atomic ground state $5S_{1/2}$ (F=2) to the Rydberg state $61S_{1/2}$, we use a two-photon transition based on counter-propagating laser beams at 780 nm and 480 nm, with a large detuning from the intermediate state $5P_{3/2}$ (Fig. 1B). A typical measurement sequence, similar to the one described in (19), consists in scanning the frequency of one of the two Rydberg lasers while monitoring the atomic fluorescence at 780nm. As we sweep across the two-photon resonance between $5S_{1/2}$ (F=2) and $61S_{1/2}$, some atoms are excited to the Rydberg state from where they can exit the trap due to ionization from the ambient blackbody field, resulting in a drop of the atom number which is maximum on resonance. By carefully adjusting the laser parameters, we obtain highly-contrasted fluorescence dips from which we can precisely extract the resonance frequencies, a technique known as trap-loss spectroscopy (19, 26-33).

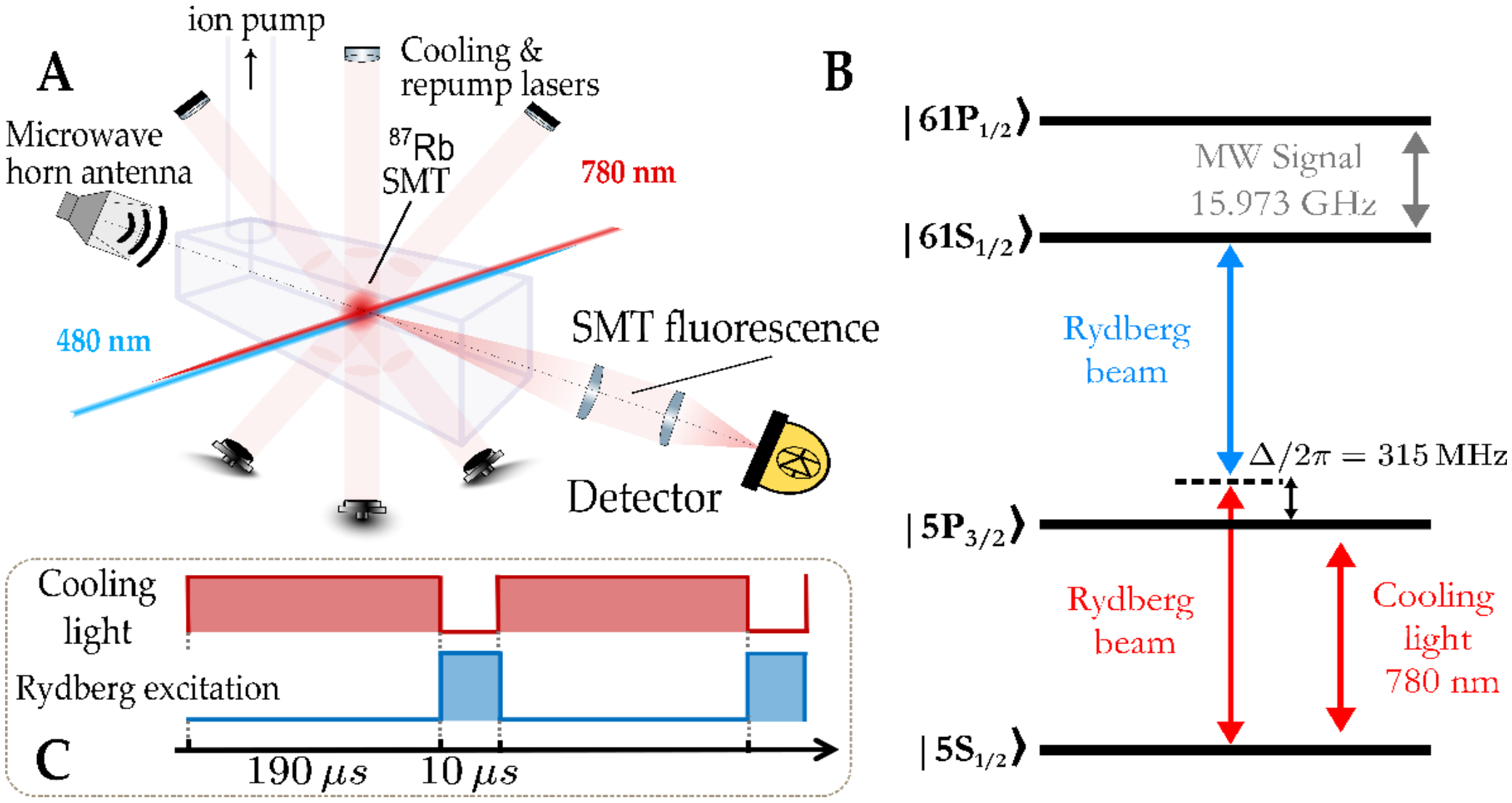

**Fig. 1. Experimental setup. A.** Sketch of the dielectric (all-optical) sensor head. A cloud of cold atoms is created in the center of a glass cell. The ion pump is connected to the main chamber by a tube to minimize perturbations to the atoms. The fluorescence from the atomic cloud is monitored as the two-photon Rydberg excitation is scanned across the $5S_{1/2}$ to $61S_{1/2}$ resonance. **B.** Main energy levels and coupling lasers involved in the experiment. The indicated value for Δ corresponds to the second series of measurement described in the main text. **C.** Experimental timing protocol for the second series of measurements, where Rydberg excitation is performed only when the cooling light is turned off. The sequence uses a 5 kHz modulation frequency with a Rydberg pulsed excitation of 10 $\mu$s, resulting in a duty cycle of 5 %.

In a first series of measurements, we probe the 2-photon transition while keeping the cooling light continuously turned on. The resulting trap-loss spectrum is shown on Fig. 2 (red curve). It exhibits two resonances, which correspond to the Autler-Townes doublet resulting from the mixing of $5S_{1/2}$ (F=2) and $5P_{3/2}$ by the cooling light. The linewidth of the fluorescence dips is on the order of 12 MHz (FWHM). It is explained by the combination of several effects resulting from the presence of the cooling light: mixing of the ground state with the intermediate state $5P_{3/2}$, whose linewidth is around 6 MHz, frequency and amplitude noise of the cooling light and spatial inhomogeneities of the latter across the atomic cloud.

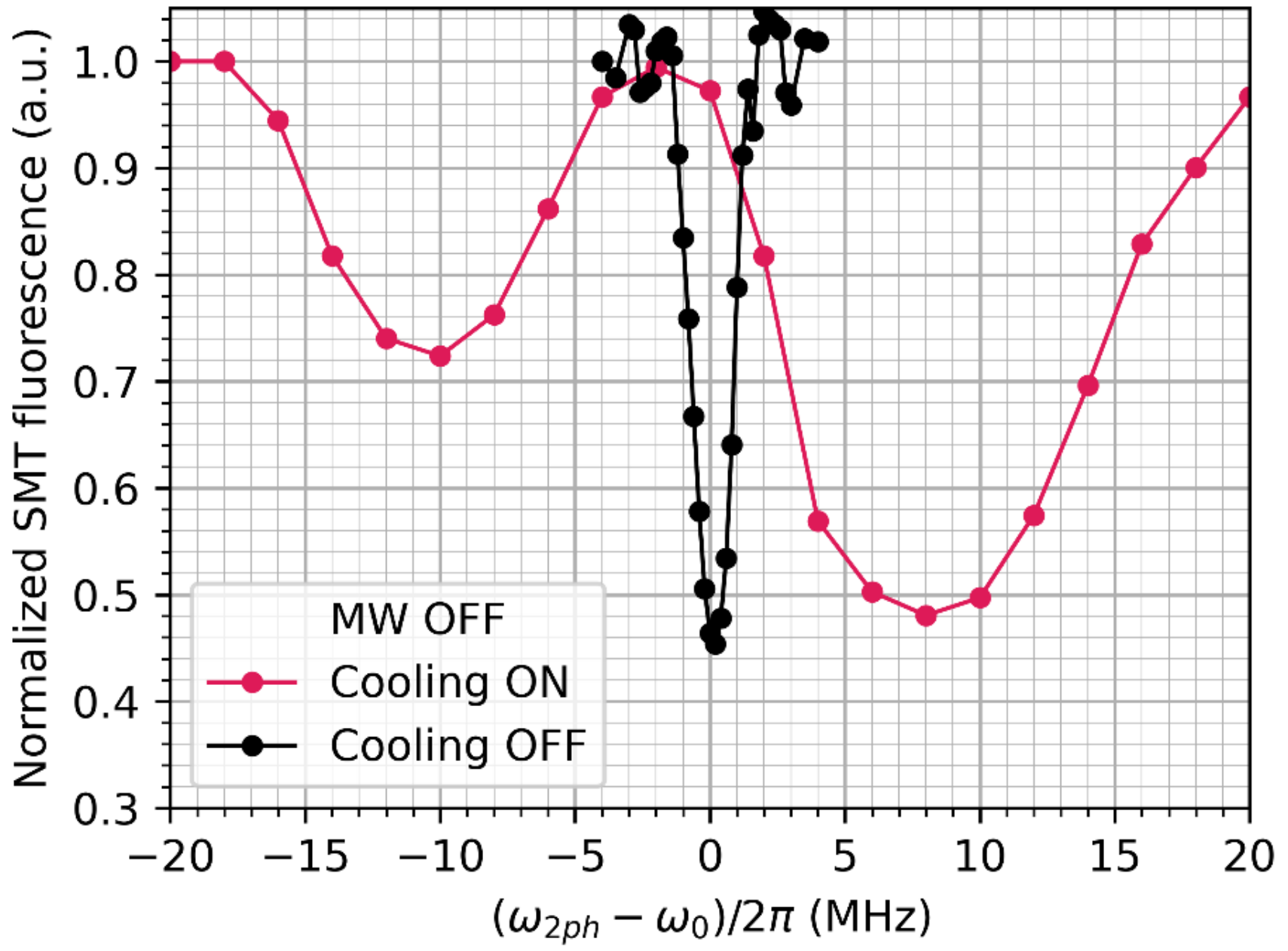

**Fig. 2. $61S_{1/2}$ Rydberg trap-loss spectroscopy with and without cooling light.** Measured fluorescence of the atoms as a function of the two-photon detuning across the ground to $61S_{1/2}$ transition. The reference $\omega_0$ corresponds to the position of the spectroscopic feature in the absence of cooling light. Each point corresponds to an averaging of the fluorescence signal during 10s, with a frequency step of 200 kHz (black dots) and 2 MHz (red dots).

In a second series of measurements, we implement a temporal sequence where Rydberg excitation is performed only when the cooling light is switched off. For this, we alternate between a cooling and trapping phase of 190 $\mu s$ and a Rydberg excitation phase of 10 μs, as illustrated on Fig. 1 C. During this sequence, the number of atoms in the SMT is barely affected by the fact that the cooling light is off 5% of the time. As in the first series of measurements, we monitor the fluorescence signal, which is averaged over several cycles by the photodetector. Because the Rydberg lasers are applied only 5% of the time, we have to adjust their intensity and intermediate detuning to increase the two-photon Rabi frequency in order to get a contrast similar to the first series of measurement. Interestingly, the reduction of the duty cycle is partly compensated by the fact that a narrower linewidth leads to a more homogeneous excitation, see the Material and methods section for more details. As we scan the two-photon frequency across the $5S_{1/2}$ (F=2) to $61S_{1/2}$ transition, we get the spectroscopic feature of Fig. 2 (black curve). Remarkably, the linewidth of the fluorescence dip is now on the order of 1.5 MHz (FWHM), almost ten times narrower than in the presence of cooling light, with no contrast reduction. This measured linewidth is however higher than the predicted Doppler linewidth for a cloud of atoms at 300 µK, on the order of 520 kHz (FWHM). We attribute this difference to a combination of saturation of the trap-loss spectroscopy curve and Zeeman broadening by the ambient magnetic field on the order of a fraction of a Gauss. Interestingly, the frequency of the bare transition measured by the black curve of Fig. 2 is off-centered by a few MHz with respect to the Autler-Townes doublet measured in the presence of cooling light. This is consistent with theory, which predicts that this difference should be equal to half the detuning between the cooling light and the $5S_{1/2}$ (F=2) to $5P_{3/2}$ transition, on the order of 3 MHz.

## RF measurements

In order to demonstrate RF sensing, we use a microwave (MW) horn to send onto the atoms a field at $\omega_{MW} \approx 2\pi \times 15.973$ GHz, on resonance with the transition frequency $\omega_0$ between the two Rydberg states $61S_{1/2}$ and $61P_{1/2}$ (see Fig. 1. B). We use the protocol described in the previous section to apply the Rydberg lasers only when the cooling light is off. As we scan the two-photon detuning of the Rydberg lasers in the vicinity of the $5S_{1/2} \leftrightarrow 61S_{1/2}$ transition, the trap-loss spectrum evolves, in the presence of the MW field, into two Autler-Townes doublets (see Fig. 3). The maximum RF power that we can send to the atoms, limited by our frequency generator, was measured to be on the order of 3 dBm at the input of the MW horn, corresponding to a maximum measured Autler-Townes splitting of about 223 MHz. Choosing for the quantization axis the direction of propagation of the MW field, the restriction of the Hamiltonian to the $\{|61S_{1/2}, m_J = \pm 1/2\rangle,\ |61P_{1/2}, m_J = \pm 1/2\rangle\}$ manifold reads in the rotating wave approximation:

$$\hat{H} = \hbar \begin{pmatrix} -\Delta_{\mathrm{MW}} & \Omega_+/2 & 0 & 0 \\ \Omega_+^*/2 & 0 & 0 & 0 \\ 0 & 0 & -\Delta_{\mathrm{MW}} & \Omega_-/2 \\ 0 & 0 & \Omega_-^*/2 & 0 \end{pmatrix}$$

where $\Delta_{\mathrm{MW}} = \omega_{MW} - \omega_0$ is the detuning of the microwave field and $\Omega_{\pm,-}$ are the Rabi frequencies of the two circular components of the microwave field. The presence of four spectral lines is the signature of an elliptical polarization ($|\Omega_+| \neq |\Omega_-|$). On resonance ($\Delta = 0$), the measured frequencies of the four spectral lines $\omega_1 < \omega_2 < \omega_3 < \omega_4$, referenced to the peak position in the absence of MW, yield the modulus of the MW electric field independently of the MW polarization using the formula:

$$\Delta\omega_{AT} = \sqrt{\frac{\omega_3^2 + \omega_4^2}{2}} + \sqrt{\frac{\omega_2^2 + \omega_1^2}{2}} = \sqrt{|\Omega_+|^2/2 + |\Omega_-|^2/2} = \frac{d.E}{\hbar} \qquad (1)$$

where $E$ is the amplitude of the microwave field and $d$ the dipole matrix element associated to a $\pi$ transition between $|61S_{1/2}, m_J = 1/2\rangle$ and $|61P_{1/2}, m_J = 1/2\rangle$, on the order of $|d| = 1291{,}5\ e.a_0$ (34) with $e$ the elementary charge and $a_0$ the Bohr radius. The value of $\Delta\omega_{AT}$ obtained from the simultaneous measurement of $\omega_1$, $\omega_2$, $\omega_3$ and $\omega_4$ is plotted on Fig. 4 as a function of the square root of the MW power, showing a linearity better than 1% over a 32 dB power range.

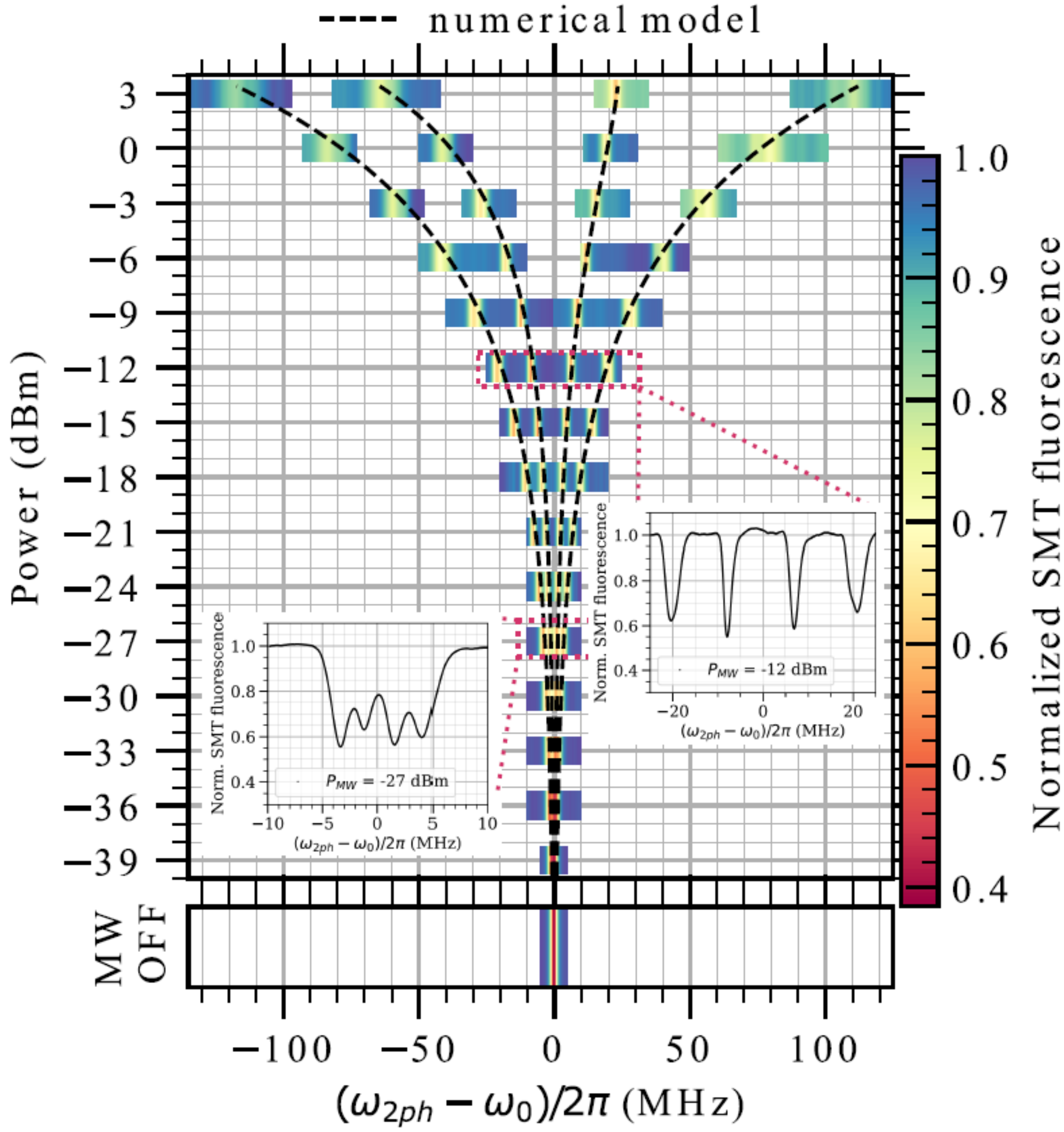

**Fig. 3. Trap-loss spectroscopy in the absence of cooling light for various levels of MW power.** The indicated MW power ($P_{MW} = -39$ to 3 dBm) was measured at the input of the MW horn. Measured fluorescence levels are plotted as a function of the two-photon detuning, $\omega_0$ accounting for the trap-loss peak position without MW. The black dashed curves result from the theoretical estimation of the peaks position, taking into account the influence of neighboring Rydberg states and our best estimate of the MW ellipticity, as described in the main text.

To quantitatively reproduce the experimental data shown on Fig. 3, it is necessary to take into account the energy shifts induced by the non-resonant coupling between $61S_{1/2}$, respectively $61P_{1/2}$, and their neighboring Rydberg states (19). Those energy shifts can be calculated theoretically as a function of the MW power, and then used to infer the absolute value of the ellipticity $\chi$ of the MW field that best fits the experimental data (dashed lines on Fig. 3), defined by $|\sin(2\chi)| = ||\Omega_+|^2 - |\Omega_-|^2|/(|\Omega_+|^2 + |\Omega_-|^2)$. This leads to the measured value $|\chi| \approx 0.735$ (see the Materials and Methods section for more details). We attribute this ellipticity, despite the fact that we send a linearly polarized field at the MW horn, to uncontrolled reflections and diffraction from different parts of the experimental setup.

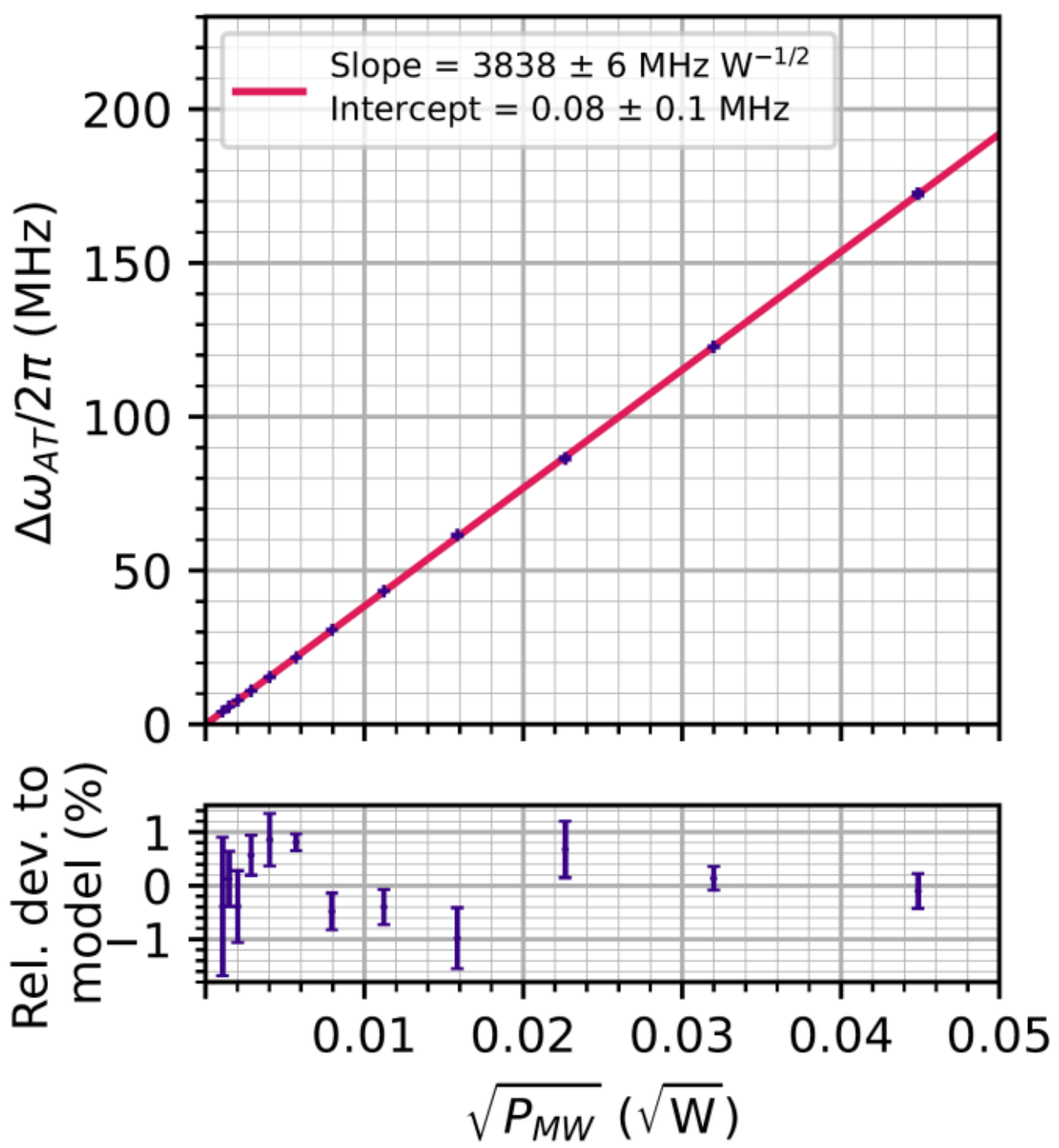

**Fig. 4. Linearity of the frequency response.** Measured splitting $\Delta\omega_{AT}/2\pi$ as a function of the square root of the applied microwave power at 15.973 GHz. The red line represents a linear fit to the data, giving an intercept of $0.08 \pm 0.10$ MHz. The error bars on the $x$-axis correspond to an estimated uncertainty of 0.01 dBm in the applied microwave power and are smaller than the marker size, while the error bars on the $y$-axis correspond to the uncertainty derived from the fits of the peak-frequencies using equation (1). The subset represents the residuals between the experimental data and the linear fit.

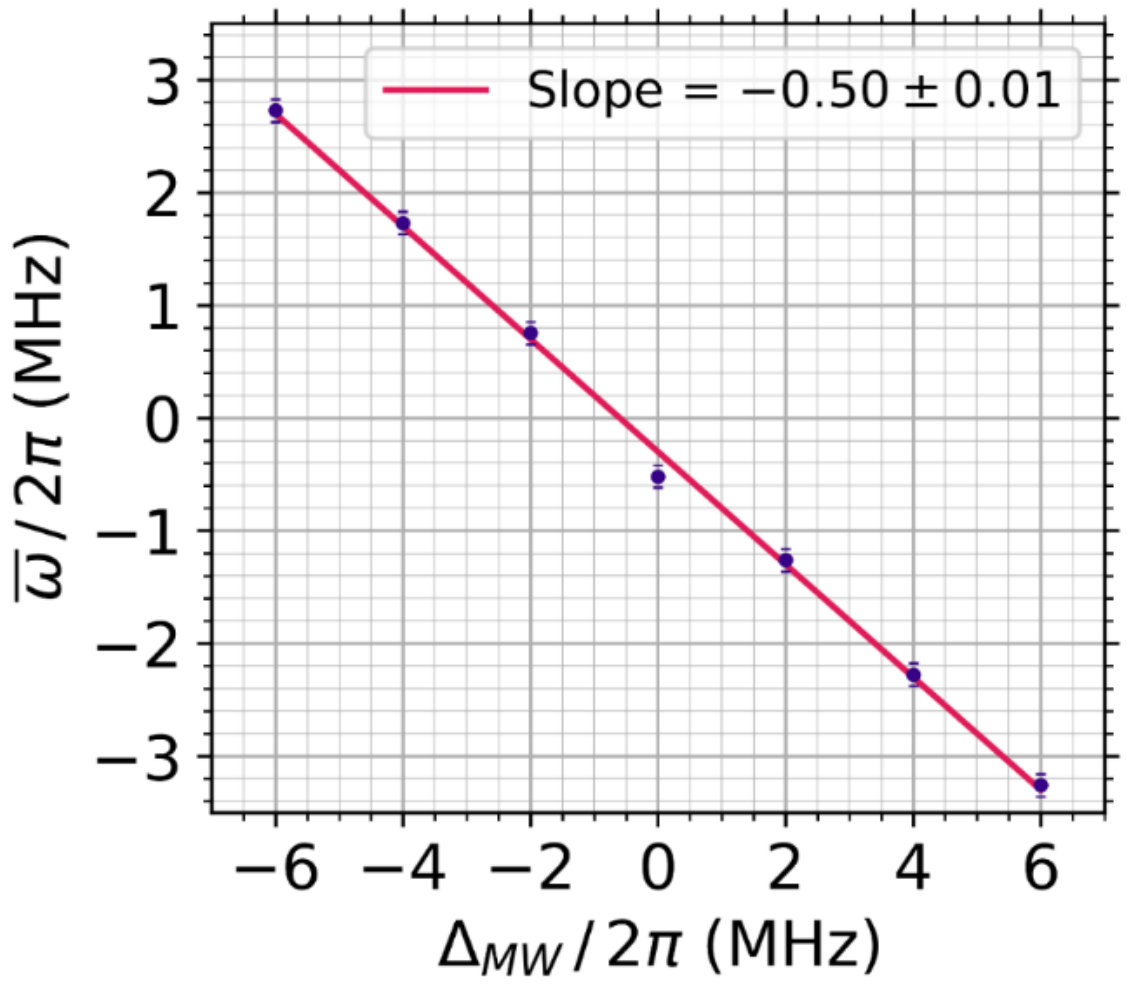

**Figure 5: Measurement of the MW frequency.** The average position of the four trap-loss spectroscopy peaks, following equation (2), is plotted as a function of the MW field detuning, showing a linear dependence with a -1/2 slope which can be used to recover the MW frequency from a single-shot measurement.

Interestingly, it is also possible to measure the detuning $\Delta_{\mathrm{MW}}$ of the MW field with respect to the bare $61S_{1/2} \leftrightarrow 61P_{1/2}$ transition by looking at the average position of the Autler-Townes doublets, defined as:

$$\varpi = \frac{\omega_1 + \omega_2 + \omega_3 + \omega_4}{4} = -\frac{\Delta_{MW}}{2} \qquad (2)$$

The linear dependence of $\varpi$ on $\Delta_{\mathrm{MW}}$ with a -1/2 slope is confirmed by the measurements shown on Fig. 5.

The width of the trap-loss spectral features, which was measured to be 1.5 MHz FWHM in the absence of applied MW field, increases linearly with the amplitude of the MW field, as shown on Fig. 6. A linear fit of the data shows a relative inhomogeneous broadening of about 10% which we attribute to the spatial inhomogeneity of electric field across the 0.5 mm-long cloud of cold atoms resulting from standing wave effects inside the cell (35), with significant variations on the length scale $\lambda/2 = c/(2f_{MW}) \simeq 10$ mm. Noticeably, this effect is much smaller than for MW sensors based on Rydberg atoms in thermal vapors, where the region of interaction between the lasers and the atoms is typically on the order of a few cm to get enough optical density.

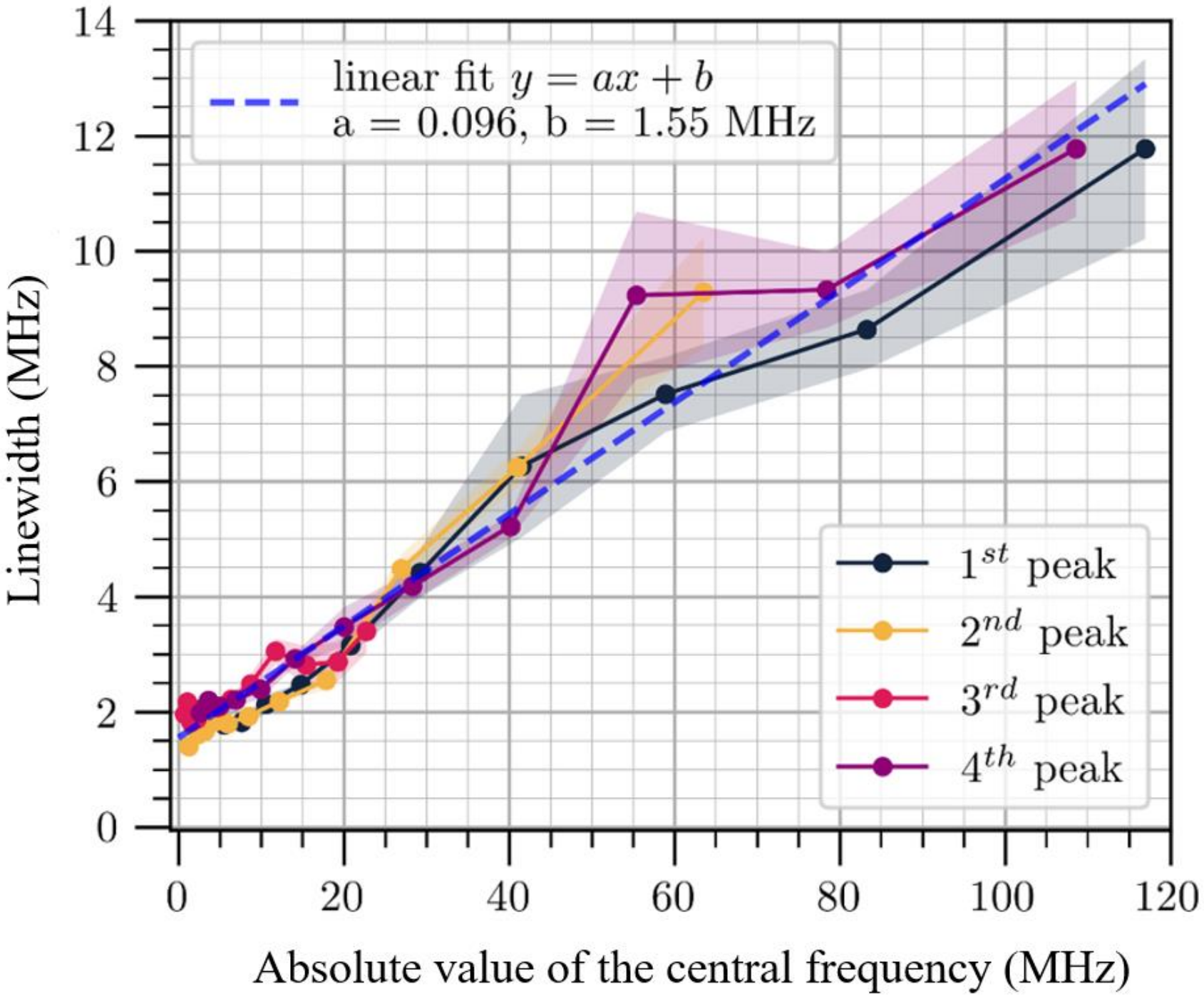

**Fig. 6. Spectral broadening for increasing MW power.** Measured linewidth (FWHM) of the trap-loss spectral features as a function of their central frequency $|\omega_i|$, referenced to the measured frequency in the absence of MW. The shaded areas represent the estimated error at $1\sigma$ from the fit. A linear fit to the data (dashed blue line) gives an intercept of $b = 1.55 \pm 0.1$ MHz and a slope of $a = 0.096 \pm 0.005$.

We also evaluated the long-term stability of our measurements. Two situations were studied: without any MW signal (Fig. 7 blue curve) and with a $-21$ dBm signal at the horn input (Fig. 7 red curve), corresponding to $\Delta\omega_{AT}/2\pi = 11$ MHz. We recorded the trap-loss signals during almost five hours for both cases and we fitted every signal to extract the frequencies of the different resonances. In the case without MW, we plotted the Allan deviation of the single peak frequency with respect to the averaging time. For small values of $\tau$, it is well approximated by 70 kHz$\sqrt{s}/\sqrt{\tau}$, revealing the presence of white noise and a short-term sensitivity of 70 kHz$/\sqrt{Hz}$, corresponding to an inferred sensitivity of 43 $\mu$V/cm$/\sqrt{Hz}$. The resonance frequency can be determined with a 6 kHz-resolution (3 $\mu$V/cm) after 300 s of averaging. At longer time scales, we observe fluctuations of this frequency limited to 9 kHz with a characteristic timescale of 2500 s. When the microwave is on, we achieve a resolution of 8 kHz for the measurement of $\Delta\omega_{AT}/2\pi$ after 3000 s, corresponding to a relative stability of $1.5 \times 10^{-3}$ on the measured MW power.

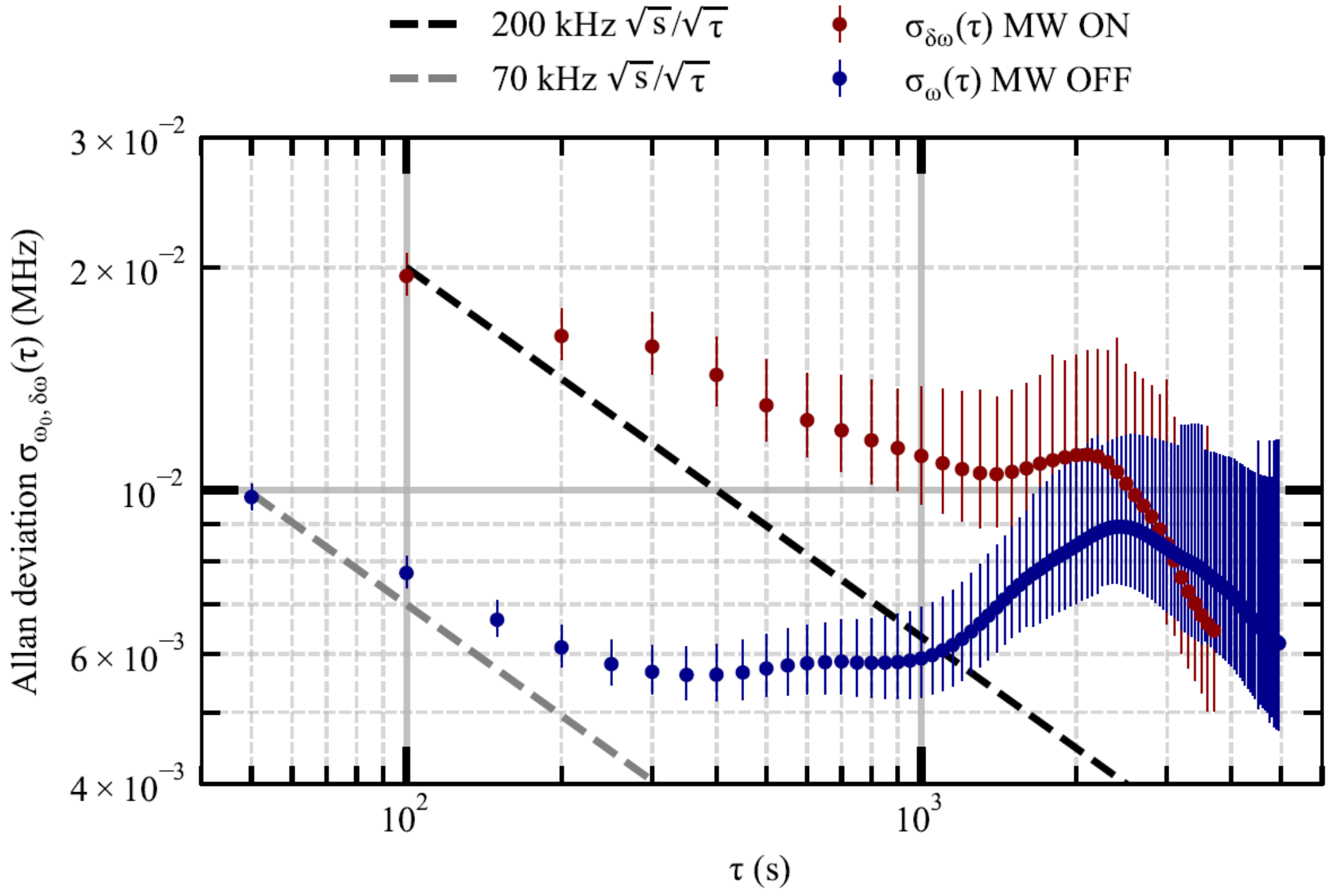

**Fig. 7. Long term stability.** Overlapping Allan deviation of trap-loss signals, in the absence of cooling light, with and without MW for an overall integration time of five hours. The MW ON case corresponds to a splitting $\Delta\omega_{AT}/2\pi = 11$ MHz. Error bars correspond to estimated confidence intervals as defined in the materials and method.

## DISCUSSION

The measurements reported in this paper demonstrate the possibility to combine the advantages of a fully optical setup with those of using cold atoms for the metrology and sensing of RF fields. The fully optical setup comes with the possibility to build a dielectric (metal-free) sensor head comprising a glass cell and optical fibers only, improving compactness and minimizing perturbations to the measured RF field. By removing the need for a magnetic field gradient to trap and cool the atoms, it furthermore allows to rapidly alternate between the trap and the Rydberg lasers, which led to a substantial improvement of the sensor performance by narrowing the linewidth of the spectroscopy peaks and removing lightshifts from the cooling lasers. The use of cold atoms instead of thermal vapors comes with several advantages in this context. First, it strongly reduces Doppler effect and transit time broadening, which are the usual limitations to the linewidth of the EIT peaks in thermal vapors. The reduction of the Doppler effect also allows to use a direct two-photon transition between the ground and the Rydberg states, with a large detuning from the intermediate state, whose linewidth is ultimately limited by the lifetime of the Rydberg state, provided other broadening mechanisms such as residual Doppler effect or magnetic field inhomogeneities are suppressed. Moreover, cold atoms allow local measurements of the MW field with a sub-millimeter spatial resolution, 40 times smaller than the wavelength for a 16 GHz field. This mitigates the broadening of the peaks and the associated loss of contrast, which usually occur for large values of the RF power. It allowed us to get highly-contrasted peaks even for an Autler-Townes splitting as large as 223 MHz. The maximum splitting is limited in our case by the total available MW power. In combination with the narrow linewidth of the peaks that we obtained at low MW power, on the order of 1.5 MHz, this leads to the unprecedented value of 43 dB for the dynamic range of the SI-traceable measurement of the MW power. Ultimately, we expect the maximum dynamic range at high field strengths to be bounded by coupling to the neighboring $61P_{3/2}$ state, which distorts the linearity of the Autler-Townes splitting as the MW Rabi frequency becomes comparable to the frequency difference between $61P_{1/2}$ and $61P_{3/2}$ (on the order of 437 MHz). This results in an inferred dynamic range of $20\log\left(\frac{437}{1.55}\right) = 49$ dB for this specific Rydberg manifold.

One limitation of the measurements reported in this paper comes from the fluctuations of the fluorescence level from the atoms in the SMT, which set the noise level at short timescale and create instabilities through thermal drifts at long timescale. We attribute these technical fluctuations, which we measure to be about 10%, to the frequency instabilities of about 1 MHz of our cooling light and to polarization fluctuations of the trapping beams. Despite those fluctuations, which could be upgraded in future experiments, we reach a short-term sensitivity of 43 $\mu$V/cm$/\sqrt{Hz}$ for a 16 GHz field, comparable to the state-of-the-art of experiments based on (SI-traceable) Autler-Townes measurements (36).

Another limitation of the current technique is the response time to a change in the MW power, on the order of a fraction of a second, limited by the loading time of the trap. This limitation is not intrinsically linked to the use of cold atoms, but rather to the trap-loss spectroscopy technique which monitors the steady-state number of atoms in the trap. Other techniques involving for example monitoring the absorption of a probe beam would lead to a faster response time, ultimately limited by the width of the spectroscopy peaks.

Future work on this experiment will involve in the short term the stabilization of the fluorescence level, which impacts the current level of performance. A further reduction of the linewidth at low MW power could also allow to resolve the Zeeman sublevels resulting from the Earth magnetic field, and select a transition insensitive to magnetic fluctuations such as $5S_{1/2}$, F=2, $m_F$=2 to $61S_{1/2}$, $m_J$=1/2. In the longer term, transitioning from trap-loss spectroscopy to absorption measurements would allow much faster sampling rates. The design of the sensor head could also be improved to minimize disturbances to the electric field amplitude and polarization. Finally, several super-molasses traps could be created in the same vacuum chamber (or used to load optical tweezers) for THz imaging (37), THz detection at the single-photon level (38), long term measurements for the detection of dark photons (39) or multiparameter measurements to detect the phase (40) or the angle-of-arrival (41) of the MW field. The metrological platform developed in this work for MW sensing also paves the way to new applications of super-molasses trap to other quantum sensors, such as clocks or gravimeters.

## MATERIALS AND METHODS

### Experimental design

The super-molasses trap that we use as an all-optical source of cold atoms has been extensively described in Reference (25). The loading rate of the trap is tuned to $0.6\ s^{-1}$ with the dispenser current. Thanks to a saturated absorption system, the repump laser is locked to the $^{87}$Rb transition between $|5S_{1/2}, F=1\rangle \rightarrow |5P_{3/2}, F'=2\rangle$. The cooling light is created thanks to an EOM controlled by a Voltage-Control Oscillator (VCO) that produces a cooling sideband 6.58 GHz lower in frequency. A 3: 3 fibered coupler splits the light into three beams for the retroreflected cooling beams along the three dimensions. We use approximately 3 mW of 780 nm laser power per cooling beam for the optical trap, with a diameter of 7.2 mm. The cooling light is red-detuned by about $\sim 1\ \Gamma$ from the cooling transition, $\Gamma = 2\pi \times 6.06$ MHz being the linewidth of the $5P_{3/2}$ state. Fluorescence is collected with two lenses and imaged onto a Si photodiode with a 3.3 MΩ gain resistor to obtain a $\sim 1$ V level signal, reducing the bandwidth to about 330Hz. To probe the $61S_{1/2}$ Rydberg state with a two-photon transition from the ground state $5S_{1/2}$, we use a RIO Planex diode at 1560 nm, which is frequency-doubled to 780 nm, and a direct amplified external cavity laser diode at 480 nm from MOGLabs. Rydberg laser beams are frequency-stabilized to an ultra-stable cavity using a fast Pound-Drever-Hall lock, resulting in linewidths below 100 Hz for both lasers. By locking a sideband generated by an EOM to a fixed cavity resonance and sending the $0^{th}$ order onto the atoms, we can tune the frequency of the latter across several hundreds of MHz at both wavelengths. We typically use Rydberg lasers beams of $\sim 0.7$ mm waist with 35 mW of optical power at 480 nm and 6 mW at 780 nm. For the first series of measurements, we set the intermediate detuning $\Delta$ to $2\pi \times 865$ MHz leading to an estimated two-photon Rabi coupling of $\Omega_{2ph} = 2\pi \times 14$ kHz at the maximum available power for both Rydberg laser beams. For the second series of measurements, we apply a Rydberg excitation pulse of 10 $\mu s$ every 200 µs, corresponding to a duty cycle of 0.05 (see Fig. 1 C). In this regime, the linewidth of the spectroscopy peak $\gamma_d$ is reduced from 12 MHz to 1.5 MHz, leading to an increase by a factor of 8 in the Rydberg excitation rate $R_{exc} = \left(\Omega_{2ph}\right)^2/2\gamma_d$ which is however counterbalanced by the fact that the excitation occurs 5% of the time only. To get a loss rate on Rydberg resonance similar to the first case, we increase the two-photon coupling strength by tuning the intermediate detuning to $\Delta = 2\pi \times 315$ MHz. Interestingly, the average fraction of atoms in the Rydberg state $R_{exc}/\Gamma_{sp}$ when the Rydberg lasers are applied is also increased by a factor 8 and is now on the order of 10%, even though we do not observe any significant sign of Rydberg-Rydberg interactions. For the RF measurements, we use a $K_u$ band adapted microwave horn antenna, which generates a linearly polarized wave constituting the RF-signal sent onto the atoms.

### Estimation of the circularity of the MW field taking into account neighboring Rydberg states

The numerical model presented Figure 3 (dashed lines) comes from the diagonalization of the Hamiltonian restricted to the $\{|61S_{1/2}, m_J = \pm 1/2\rangle, |61P_{1/2}, m_J = \pm 1/2\rangle\}$ manifold (dimension 4), including the shifts resulting from the non-resonant coupling with the neighboring Rydberg states $61P_{3/2}, 60P_{1/2}, 60P_{3/2}, 62S_{1/2}, 60D_{3/2}$ and $59D_{3/2}$ (19). These shifts are estimated using the perturbation theory. They take the general form $|\Omega_{ij}|^2/(4\Delta_{ij})$, where $\Omega_{ij}$ is the Rabi coupling taking into account the polarization of the MW field, and $\Delta_{ij}$ the MW detuning with respect to the $i \leftrightarrow j$ transition. Here, we neglect the contribution of neighboring atomic levels whose energy shift was less than 100 kHz for a reference splitting of $\Delta\omega_{AT}/2\pi = 100$ MHz. To obtain the full Hamiltonian, we add to the matrix $\hat{H}$ given in the main text the diagonal terms corresponding to the shifts from the non-resonant couplings. We diagonalize this 4x4 matrix, with the absolute value of the MW ellipticity $|\chi|$ being the only free parameter. Our best estimate of $|\chi|$ is then obtained by minimizing the distance between the measured frequencies and their theoretical values $\sum_i (\omega_{exp,i} - \omega_{theory,i})^2$ for *i* spanning the 4 eigenfrequencies and 12 different power values between -30 dBm and +3 dBm.

**Statistical Analysis**

We acquire the data using a 16-bit resolution and a 101Hz sampling frequency. Except for Figure 2 where the fluorescence is recorded with discrete frequency steps, we acquire the spectra by scanning the two-photon frequency of the Rydberg lasers at a rate of 0.4 MHz/s, alternating between both directions, to mitigate the distortion of spectral features by dynamical effects (19).

For Figures 3 to 6, we average the signal over ten ascending (respectively descending) sweeps. For each scan direction, we fit the averaged spectral profiles with asymmetric Gaussian functions $y(x) = y_0 + A \cdot e^{-\frac{(x-x_0)^2}{2(w+a(x-x_0))^2}}$. The estimated parameters (central frequency and linewidth) and their $1\sigma$ errors are then obtained by averaging between the ascending and the descending values from the fit. We measure the uncertainty in the applied MW power to be 0.01 dBm using a spectrum analyzer.

For Figure 7, we fit the single spectra instead of the averaged curves. We then plot the overlapping Allan deviation of the central frequency (respectively the Autler-Townes splitting) averaged within pairs of consecutive ascending and descending scans. The error bars associated to each averaging times correspond to confidence intervals calculated using the function '*confidence_interval*' from the '*AllanTools*' Python library. For the estimation of these intervals, we compute the equivalent degrees of freedom from the Greenhall algorithm (42), implemented in the '*allantools.edf_greenhall*' function, where we estimate the noise type of the $\omega_0(t)$ (resonance frequency) and $\delta\omega(t)$ (AT-splitting) dataset (*e.g.* white noise, frequency drift and frequency flicker noise) using autocorrelation functions at different integration times (43).

## ACKNOWLEDGEMENTS

We acknowledge support from Olivier Le Traon, Andrei Dragomir, Jonathan Woods and Jean-Michel Leandri.

## FUNDING

We acknowledge funding by the CW ITP program (CACQTUS project). This project also benefited from a French government grant managed by the Agence Nationale de la Recherche in the France 2030 framework, with the reference ANR-23-PETQ-0004, and from the project ANR-22-CE47-0009-03. Romain Granier acknowledges a PhD fellowship from Ecole Normale Supérieure Paris-Saclay in the CDSN program. Anthony El Bekai acknowledges a PhD fellowship from the graduate school of Université Paris-Saclay.

## AUTHOR CONTRIBUTIONS

Conceptualization and supervision of the research: A. Bo., S. S.
Design of the experimental apparatus: V. A., C. C., F. C., K. K., M. H., A. J., A. Bo., S. S.
Building of the experimental setup: R. G., A. E. B., M. A. C. M., C. B., V. A., C. C., F. C., K. K., M. H., A. J., A. Bo., S. S.
Methodology: R. G., N. Z., Y. B., A. Br., A. Bo., S. S.
Conducting the experiment and collecting data: R. G., A. E. B., M. A. C. M., V. A., C. C., A. Bo., S. S.
Data analysis: R. G., A. E. B., M. A.C. M., V. A., C. C., M. H., A. J., A. Bo., S. S.
Writing – original draft: R. G, A. Bo., S. S.
Writing – review and editing: all the authors

## COMPETING INTERESTS

Aquark Technologies is associated with intellectual property relating to the super-molasses trap described in this work (patents GB2638701 and GB2595746). Authors Vilius Atkočius, Chester Camm, Florence Concepcion, Konstantinos Karakostas, Alexander Jantzen and Matt Himsworth are employees of Aquark Technologies, which develops quantum technologies related to the subject matter of this study. The remaining authors declare no competing interests.

## DATA AVAILABILITY

All data used in the analyses can be made available upon reasonable request.